%% file: ms.tex
\documentclass[conference]{IEEEtran}
\IEEEoverridecommandlockouts
\usepackage{cite}
\usepackage{amsmath,amssymb,amsfonts}
\usepackage{graphicx}
\usepackage{textcomp}
\usepackage{xcolor}

\usepackage{fancyvrb}
\usepackage{subfig}
\usepackage[pass]{geometry}
\usepackage{fancyhdr}
\usepackage[normalem]{ulem}
\usepackage[hyphens]{url}
\usepackage{hyperref}
\usepackage{flushend}
\usepackage{comment}
\usepackage{threeparttable}
\usepackage{algpseudocode}
\usepackage{algorithm}
\usepackage{multicol}
\usepackage{multirow}

\def\BibTeX{{\rm B\kern-.05em{\sc i\kern-.025em b}\kern-.08em
    T\kern-.1667em\lower.7ex\hbox{E}\kern-.125emX}}
\begin{document}

\title{Exploring Modern GPU Memory System Design Challenges through Accurate Modeling}

\author{
    \IEEEauthorblockN{Mahmoud Khairy\IEEEauthorrefmark{1}, Jain Akshay\IEEEauthorrefmark{1}, Tor Aamodt\IEEEauthorrefmark{2}, Timothy G. Rogers \IEEEauthorrefmark{1}}
    \IEEEauthorblockA{\IEEEauthorrefmark{1}Electrical and Computer Engineering, Purdue University
    \\\{abdallm,akshayj,timrogers\}@purdue.edu}
    \IEEEauthorblockA{\IEEEauthorrefmark{2}Electrical and Computer Engineering, University of British Columbia
    \\\ aamodt@ece.ubc.ca}
}

\maketitle
\thispagestyle{plain}
\pagestyle{plain}
\begin{abstract}
This paper explores the impact of simulator accuracy on architecture design decisions
in the general-purpose graphics processing unit (GPGPU) space.
We perform a detailed, quantitative analysis of the most popular
publicly available GPU simulator, GPGPU-Sim, against our enhanced version of the simulator,
updated to model the memory system of modern GPUs in more detail.
Our enhanced GPU model is able to describe the NVIDIA Volta architecture
in sufficient detail to reduce error in memory system even counters
by as much as $66\times$. The reduced error in the memory system
further reduces execution time error versus real hardware by $2.5\times$.
To demonstrate the accuracy of our enhanced model against a real machine,
we perform a counter-by-counter validation against an NVIDIA TITAN V Volta GPU,
demonstrating the relative accuracy of the new simulator versus the publicly
available model.

We go on to demonstrate that the simpler model discounts the importance of
advanced memory system designs such as out-of-order memory access scheduling,
while overstating the impact of more heavily researched areas like L1 cache bypassing.
Our results demonstrate that it is important for the academic community
to enhance the level of detail in architecture simulators as system
complexity continues to grow.
As part of this detailed correlation and modeling effort, we developed
a new {\em Correlator} toolset that includes a consolidation of
applications from a variety of popular GPGPU benchmark suites, designed to run
in reasonable simulation times. The Correlator also includes a database of
hardware profiling results for all these applications on
NVIDIA cards ranging from Fermi to Volta and a toolchain
that enables users to gather correlation statistics and
create detailed counter-by-counter hardware correlation plots with minimal effort.

\end{abstract}

\input{introduction}
\input{background}
\input{methodology}
\input{demystifying}
\input{correlation}
\input{motivation}


\input{related}

\input{conclusion}
\bibliographystyle{ieeetr}
\bibliography{all} 

\end{document}

%% file: introduction.tex
\section{Introduction}

\begin{figure}
    \centering
    \includegraphics[width=85mm]{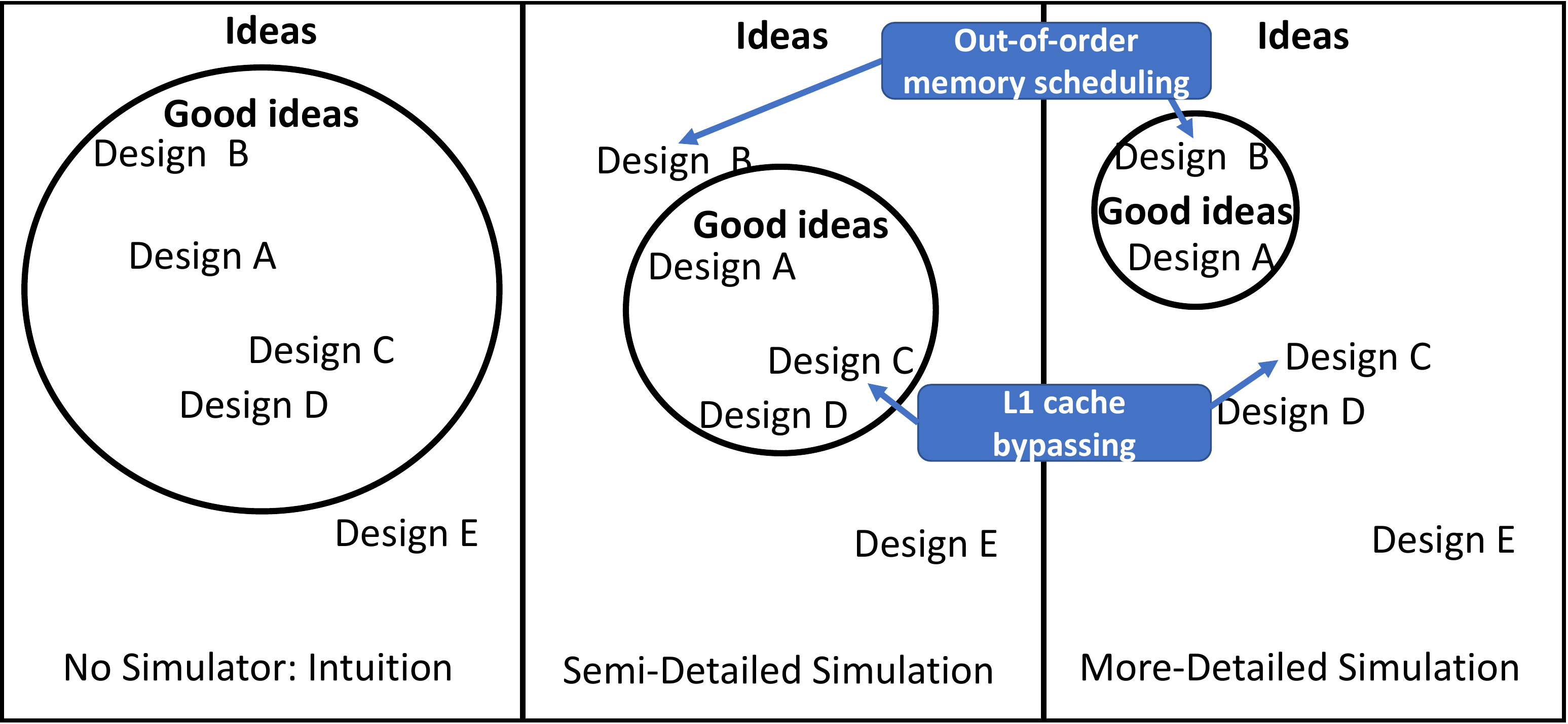}
    \caption{The architecture design and simulation process represented
    as a collection of good and ineffective ideas. As the level of simulation detail increases,
    the space of effective ideas shrinks and potentially moves.\label{fig:ideas} }
\end{figure}

In contemporary computer architecture research, simulation is commonly used to estimate
the effectiveness of a new architectural design idea. 
High-level simulators enable architects to rapidly evaluate ideas at the expense of less accurate simulation results.
Ideas that do not show promise in simulation are discarded while those that do show promise
are refined in an iterative process~\cite{CA:AQA}.
As Figure~\ref{fig:ideas} illustrates, simulation inaccuracy can
lead to the retention of design proposals with overestimated benefits,
or the rejection of design proposals with underestimated benefits.   
In one scenario promising ideas 
are throw out prematurely leading to less optimal solutions.
In the other scenario, ineffective ideas are retained longer than necessary,
leading to wasted time and effort during the architecture design process.

Our paper focuses on the simulation of massively parallel architectures,
in particular GPUs. GPUs have witnessed rapid change and a widespread increase in their
adoption with the rise of GPGPU computing and machine learning.
In academia, the design of programmable accelerators
is mostly carried out through modeling new techniques in
high level GPU simulators.
Over the past four years, there have been approximately 20 papers
per year focusing on GPU-design at the top architecture conferences.
80\% of those papers have used today's most
popular open-source GPU simulator, GPGPU-Sim~\cite{gpgpusimmanual}.
The relative popularity of GPGPU-Sim can be attributed to
several factors, but it's most appealing
aspect is perhaps the accuracy with which it models
modern GPUs (relative to other open-source solutions).
Such accuracy should provide a solid basleine for studying
important architectural ideas that are relevant to
future machines.

Recent work on validating GPGPU-Sim~\cite{jain2018quantitative}
has demonstrated that there are several areas where a lack of
detail in the performance model creates a a major source of error.
However, the bulk of the error comes from the modeling of the memory system.
Building on that observation, we perform a module-by-module
redesign of the GPU's memory system, demonstrating its improved correlation with real hardware.
Our improved performance model decreases memory system error on several
key metrics over a diverse set of benchmarks by as much as $66\times$,
resulting in a $2.5\times$ reduction in execution cycle error when modeling
and NVIDIA TITAN V GPU.
Table~\ref{ta:corr} presents the average absolute error rates of the publicly
available GPGPU-Sim 3.x versus our new model~\footnote{Simulator changes described
in this paper will be released publicly}.
The GPGPU-Sim 3.x model, or {\em old model} is a faithful representation of how papers
currently scale GPGPU-Sim to update their baseline GPU, without modifying
the code. The {\em new model} represents all the changes implemented in this paper.

\begin{table}
\small
\centering
\caption{Average absolute error and correlation rates of the publicly available
GPGPU-Sim model versus our enhanced memory system model when modeling an 
NVIDIA Volta TITAN V.}
\label{ta:corr}
\begin{tabular}{|l|l|l|l|l|}
\hline
\multirow{2}{*}{Statistic} & \multicolumn{2}{l|}{Mean Abs Error} & \multicolumn{2}{l|}{Correlation} \\ \cline{2-5} 
 & \begin{tabular}{@{}c@{}}Old\\Model \end{tabular} & \begin{tabular}{@{}c@{}}New\\ Model\end{tabular} & \begin{tabular}{@{}c@{}}Old\\Model \end{tabular} &\begin{tabular}{@{}c@{}}New\\ Model\end{tabular} \\ \hline
L1 Reqs & 48\% & 0.5\% & 92\% & 100\%\\ \hline
L1 Hit Ratio & 41\% & 18\% & 89\% & 93\%\\ \hline
L2 Reads & 66\% & 1\% & 49\% & 94\%\\ \hline
L2 Writes & 56\% & 1\% & 99\% & 100\%\\ \hline
L2 Read Hits & 80\% & 15\% & 68\% & 81\%\\ \hline
DRAM Reads & 89\% & 11\% & 60\% & 95\%\\ \hline
Execution Cycles & 68\% & 27\% & 71\% & 96\%\\ \hline
\end{tabular}
\end{table}

To demonstrate that our new model  more closely matches
the hardware, we perform a counter-by-counter validation of the improved memory system
versus the current 3.x memory system. During the course of this analysis
we uncover a number of interesting insights into how contemporary hardware works.
We utilized every publicly available resource to construct the new memory system
model, capturing the details of what others has either disclosed or
discovered~\cite{wong2010demystifying,nugteren2014detailed,mei2017dissecting,voltacitadel}.
However, in the process of correlating the memory system we also discovered several
new, previously undocumented features.
For example, we uncovered more detailed elements of the Volta streaming L1 cache
behaviour, implemented Volta's adaptive L1 cache configuration policy,
discovered that both the L1 and L2 caches are sectored, explored the fine details of the Volta
memory coalescer and finally discovered and implemented the L2 writepolicy for GPU caches
which attempt to conserve memory bandwidth in the presence of sub-sector reads.


Using this more detailed model, we perform a case-study on
the effects error in the memory system can have on
architectural design decisions.
As expected, we demonstrating that some
ideas are rendered both more and less effective
using a more detailed model.
In particular, we demonstrate that out-of-order memory access 
scheduling, which appears relatively ineffective under the old model,
yields a significant performance improvement when the level of detail
in the memory system is increased to more closely match hardware.
We then go on to demonstrate that the L1 cache throughput bottleneck
present in the current version of the simulator no longer exists in
contemporary hardware.
The introduction of streaming, banked, sectored L1s with
an allocate-on-fill policy removes a key source of contention in
GPGPU-Sim's L1 cache. Consequentially, any design aimed at mitigating
this L1 throughput bottleneck will is less effective on a contemporary
GPU, which is reflected in our enhanced simulator.

This paper makes the following contributions:
\begin{itemize}
\item It presents the most accurate open-source
GPU performance model to date, reducing the error in Volta
execution time by a factor of $2.5\times$, by reducing
error between silicon performance counters and GPGPU-Sim by $66\times$.
\item It demonstrates the effect of modeling error on architectural
design decisions made using the simulator. We concretely
demonstrate that some problems observed using the less detailed
memory model are no longer an issue in state-of-the-art machines, which 
have designed around issues like the L1 cache bottleneck but are more 
sensitive to issues like memory access scheduling.
\item It identifies and confirms a number of architectural design 
decisions that have be made in contemporary accelerators.
We discuss the potential reasons why these decisions
were made and suggest opportunities to exploit those
decisions in future research.
\item It introduces a new open-source correlator toolset that includes a simple workflow
for compiling over the GPU applications from 8 popular GPGPU benchmark suites
for CUDA versions 4.2 through 10.0, with inputs curbed for simulation.
The toolset also includes a database of real 
hardware timing and performance counters for these workloads for five different GPU
product generations and a correlation tool that allows users of
GPGPU-Sim to easily generate counter-by-counter correlation plots for
their simulator changes or their new workloads .

\end{itemize}


%% file: background.tex

%% file: methodology.tex
\section{Experimental Setup}
\label{sec:exp}

\begin{figure}
    \centering
    \includegraphics[width=85mm]{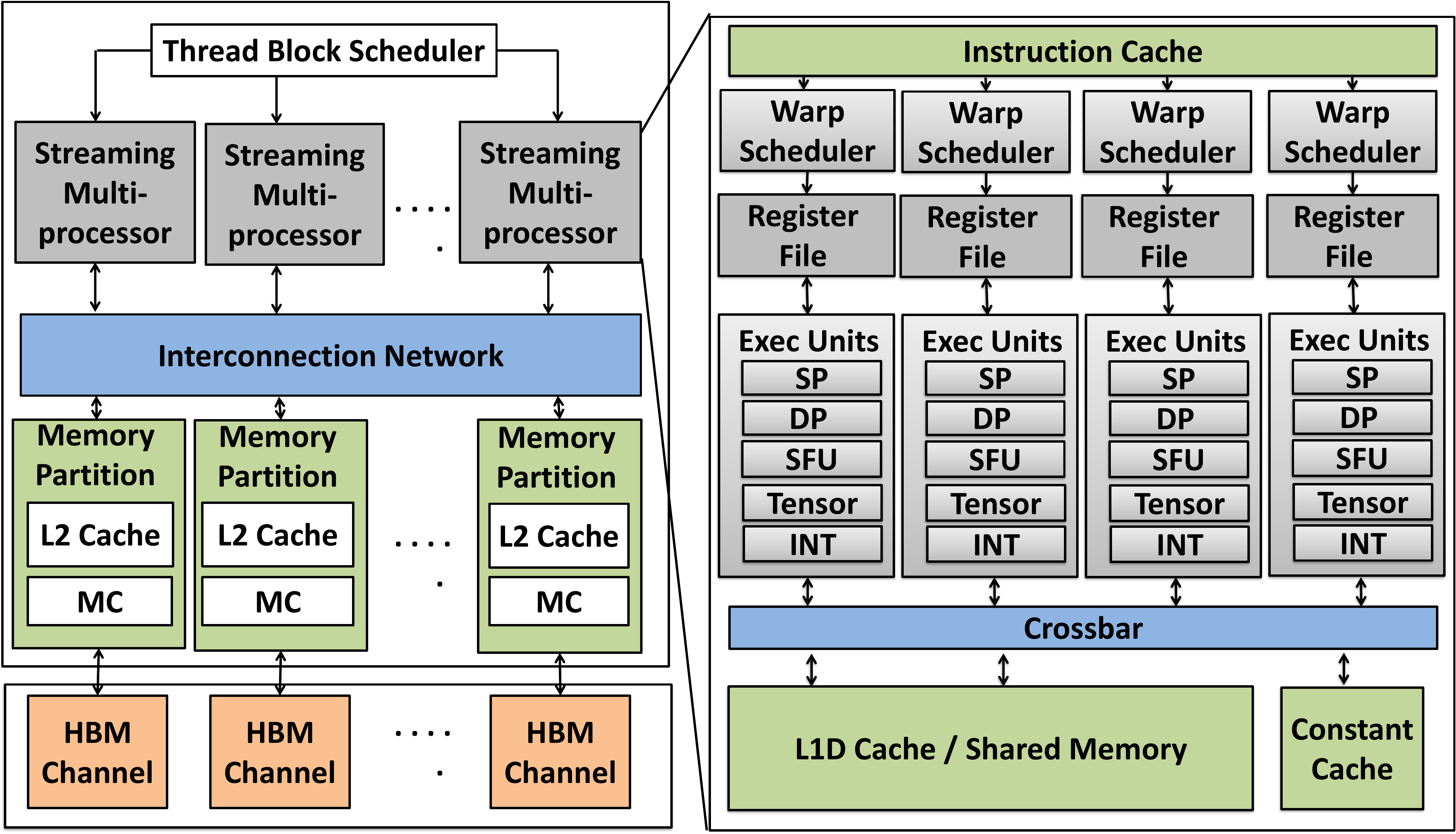}
    \caption{Volta Architecture\label{fig:volta-arch}}.
\end{figure}

We modified the GPGPU-Sim 3.x~\cite{bakhoda2009analyzing} extensively to model the new Volta architecture \cite{v100,v100micro}.
Figure~\ref{fig:volta-arch} depicts the Volta V100 architecture.
The new streaming multiprocessor (SM) is composed of four warp schedulers,
compared to two warp schedulers in Fermi architecture. Each warp scheduler has a dedicated register file
(RF) and its own execution unit (EU), including: single-precision floating point unit (SP),
double-precision unit (D), special-function unit (SFU),
integer unit (INT), and tensor core unit (TENSOR) to
accelerate 4x4 matrix multiplication \cite{v100}. A warp scheduler with its dedicated RF and EUs
is named sub-core \cite{v100micro}. All the four warp schedulers share the memory unit that
contains a memory coalescer unit and a unified cache.
The coalescer unit coalesces warp requests at the granularity of eight threads whether
the request is cached or non-cached. The schedulers and memory units are connected
via a fair round-robin crossbar. 

In Volta, the L1 cache and shared memory are combined together in a 128KB unified cache.
The shared memory capacity can be set to 0, 8, 16, 32, 64 or 96 KB and the remaining cache
serves as L1 data cache (minimum L1 cache capacity = 32KB).
Unlike previous generations, the driver adaptively configures the shared memory capacity for each kernel~\cite{nvidia2010programming}
such that it avoids shared memory occupancy bottlenecks. On the other hand, if a kernel does
not utilize shared memory, the whole unified cache (128KB) will be assigned to L1 cache.
The Volta V100 comes with High Bandwidth Memory (HBM) modules~\cite{o2014highlights,hbmstandard2013high}.
Each HBM memory module contains eight channels.
Memory channels, including the memory-side L2 cache, are connected with SMs throughout an on-chip crossbar.

\begin{table*}
  \begin{center}
    \scriptsize
   \caption{Volta V100 TITANV Configuration} \label{table:volta-config}

    \begin{tabular}{|l|c|c|}
        \hline
         & Old Model & New Model \\\hline
        \#SMs & 80 & 80\\\hline
        \#SM Configuration & \begin{tabular}{@{}c@{}}Warps per SM = 64, \#Schedulers per SM = 4, \\ \#Warps per Sched = 16, RF per SM = 64 KB \end{tabular} & \textbf{+Volta model (similar to fig\ref{fig:volta-arch} )} \\\hline
        \#Exec Units & 4 SPs, 4 SFUs & \textbf{+ 4 DPs, 4 INTs} \\\hline
        Memory Unit & Fermi coalescer (32 threads coalescer) & \textbf{Volta coalescer (8 threads coalescer) + Fair memory issue}\\\hline
        Shared Memory  & Programmable-specified up to 96 KB & \textbf{Adaptive (up to 96 KB)} \\\hline
        L1 cache & \begin{tabular}{@{}c@{}}32KB, 128B line, 4 ways, write-evict, turn-on-by-default \end{tabular}& \begin{tabular}{@{}c@{}} \textbf{+32B sector, adaptive cache (up to 128 KB),} \\ \textbf{Streaming cache, banked, latency = 28 cycles} \end{tabular}
\\\hline
        L2 cache &\begin{tabular}{@{}c@{}}  4.5 MB, 24 banks, 128B line, 32 ways, Write Back, \\ Naive write allocate, LRU, latency=100 cycles \end{tabular} &   \textbf{+32B sector, Lazy\_Fetch\_on\_Read, memory copy engine model},   \\\hline
        Interconnection & 40x24 crossbar, 32B flit  & \textbf{+advanced memory partition indexing (reduce partition camping)} \\\hline
        Memory Model & GDDR5 model, BW=652 GB/sec, latency=100ns & \textbf{HBM Model, dual-bus interface, Read/Write buffers, advanced bank indexing
}\\\hline
    \end{tabular}
  \end{center}
\end{table*}

The public version of GPGPU-sim models the now 9 year old Fermi architecture~\cite{Fermi}.
The standard practice in GPGPU-Sim to model modern GPUs, such as Volta, is to scale the configuration
of the Fermi model by increasing parameters like the number of cores, memory channels, clocks, number of warp schedulers etc.
without performing a rigorous validation of such a configuration against modern hardware.
In this work, we made material changes to the GPGPU-sim simulator code in order to accurately model
the Volta architecture. Table~\ref{table:volta-config} details the high-level architecture of the
Volta V100 TITANV~\cite{titanv} architecture we model and validate against. The left column of the table
contains the configuration of the GPGPU-sim 3.X old model obtained by scaling the Fermi resources similar
to standard practice. The right column lists the new changes, shown in bold, made to model Volta
architecture accurately. Note that, these new changes cannot be configured in the old model,
instead it requires significant code changes. All the data shown in the table are adopted from
either publicly available documents and patents from Nvidia \cite{v100,v100micro}, or via running
extensive in-house micro-benchmarks. Section~\ref{sec:cache-demy} discusses in more detail th
micro-benchmarks and the details they uncover.
The L1 cache's  latency, indexing, number of sets and line size is adopted from ~\cite{voltacitadel}.

We implement a throughput-oriented, banked, streaming, sectored L1 cache, a sectored L2 cache with a lazy-fetch-on-read
policy that is highly correlated with the real hardware. 
Further, we model CPU-GPU memory copy engine which we found is an important factor on L2 accesses and hit rate,
since all DRAM accesses go through the L2, including CPU-GPU memory copies~\cite{keplercache}.
In order to reduce uneven accesses across memory partitions~\cite{aji2011bounding}, we add an advanced partition
indexing that xors the L2 channel bits with randomly selected bits from the higher row
and lower bank bits~\cite{liu2018get}. In the memory system, we accurately model HBM.
This includes the dual-bus interface and the new per-bank refresh command~\cite{o2014highlights,hbmstandard2013high}.
Further, we implement the well-known memory optimization techniques, bank indexing and separate read/write buffers, to reduce bank conflicts and read-write inference~\cite{stuecheli2010virtual,stuecheli2010virtual,lee2010dram,chatterjee2014managing}.

For Validation, we use a wide range of applications to ensure our new model generalizes to
a variety of workloads. We indiscriminately include all the workloads we could get to run in simulation
from
the original GPGPU-Sim paper~\cite{bakhoda2009analyzing},
Rodinia 3.1~\cite{industry30},
the CUDA SDK~\cite{cudasdk},
Parboil~\cite{industry32},
Pannotia~\cite{che2013pannotia},
Lonestar \cite{burtscher2012quantitative},
SHOC~\cite{industry29},
and a collection of DOE proxy apps~\cite{lulesch}. The sum total of which is over 1400
unique kernel launches.
All the workloads are compiled with CUDA 10 and the compute capability of Volta architecture (=sm\_70). 
We updated the simulator to execute applications compiled with sm\_70 PTX.

%% file: demystifying.tex
\section{Exploring the Volta Memory System}
\label{sec:cache-demy}
The memory system in modern GPUs has changed significantly
since the last detailed modeling effort in GPGPU-Sim, which
took place for the Fermi architecture.
Unlike in CPUs, the caches in GPUs are not primarily used to reduce 
single-thread memory latency, but rather
are designed as a bandwidth filter, reducing accesses to lower
levels of memory.
Workloads with a higher hit rate in the L2 cache service more requests at the faster throughput of the L2 cache,
which is typically 2X the {{DRAM}} throughput. There have been numerous works tried to demystify GPU architecture and memory system details through micro-benchmarks for Tesla GT200~\cite{wong2010demystifying}, Fermi~\cite{nugteren2014detailed}, Maxwell~\cite{mei2017dissecting}, and recently Volta~\cite{voltacitadel}. They demystified the L1/L2 cache's  associativity, replacement policy, indexing and capacity. However, none of these works dissect the details of Volta L1 streaming cache behaviour, adaptive cache configuration, sectoring L1/L2 caches, Volta coalescer and L2 write policy, taking into account that these details have a substantial impact on the GPU Modeling. In the following subsections, we explore the design of the Volta L1/L2 cache policy in order to accurately model the Volta memory system.

\begin{figure}
    \centering
    \includegraphics[width=85mm]{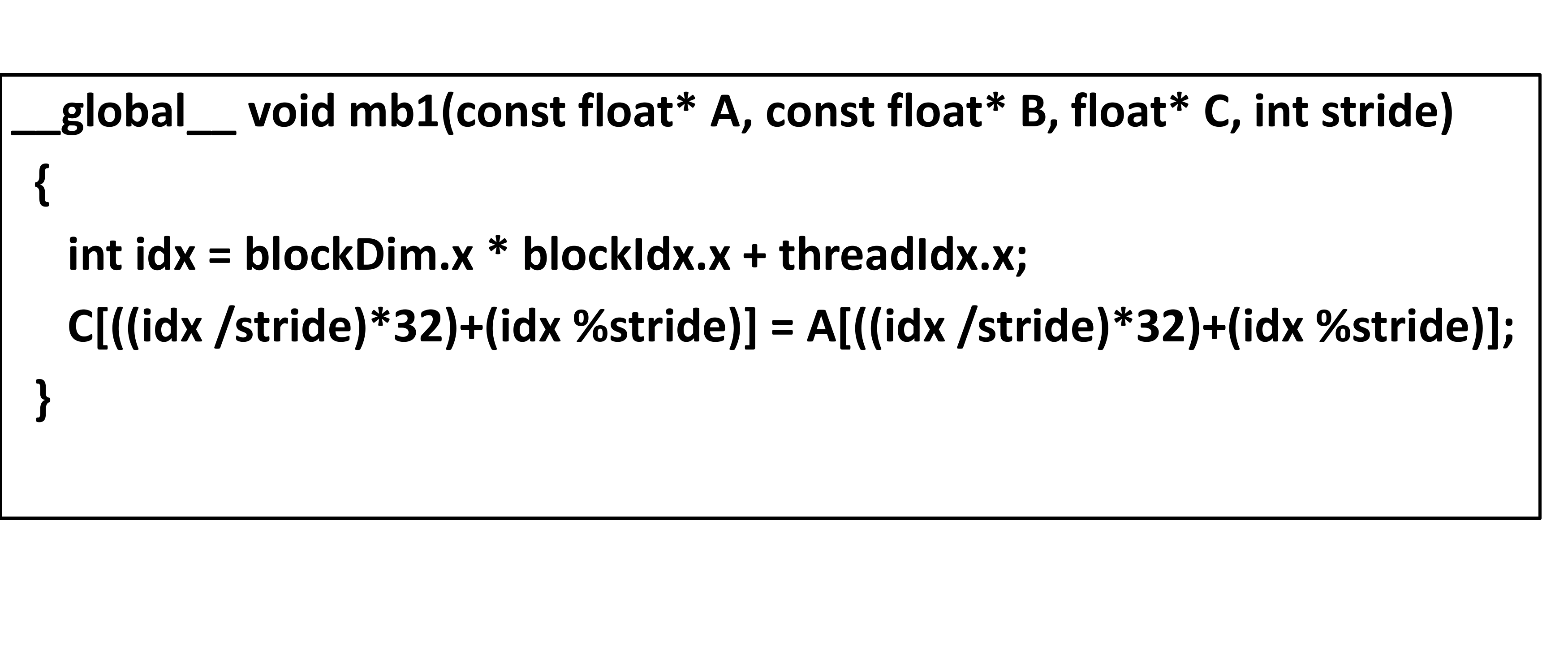}
    \caption{Volta coalescer micro-benchmark \label{fig:demystiying_coal}.}
\end{figure}

\begin{figure}
    \centering
    \includegraphics[width=85mm]{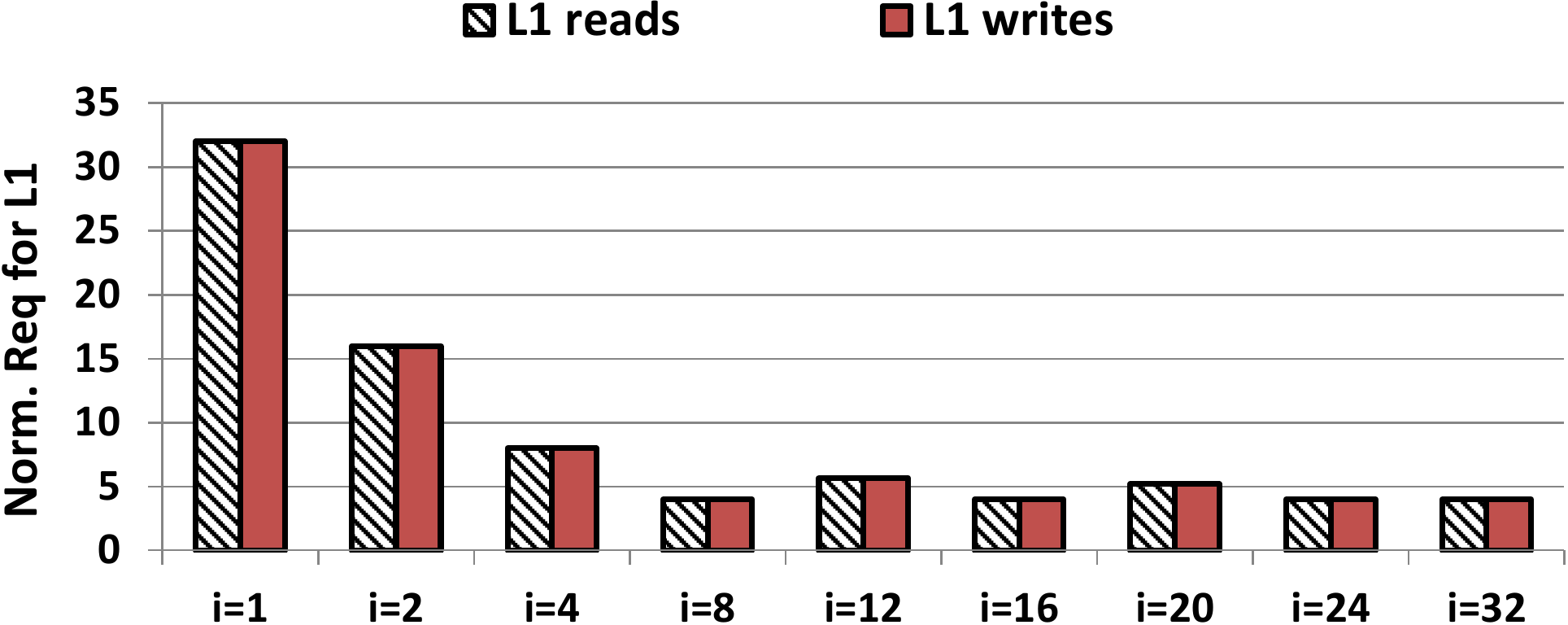}
    \caption{Volta coalescer micro-benchmark results\label{fig:coal_strides}.}
\end{figure}

\subsection{Volta Memory Coalescer \label{sec:pascal-mem}}
 
To understand the behaviour of the Volta global access coalescer,
we create the micro-benchmark shown in Figure~\ref{fig:demystiying_coal}.
We run our micro-benchmark on the NVIDIA
TITANV~\cite{titanv} and collect the
results via nvprof 10.0~\cite{nvprof}.
In the micro-benchmark, we adjust the number
of cache lines accessed by a warp based on variable {\em stride}.
When stride=1, each thread will access a different cache line,
when stride=8, each consecutive group of eight threads will access the same cache line, and so on.

\begin{figure}
    \centering
    \includegraphics[width=85mm]{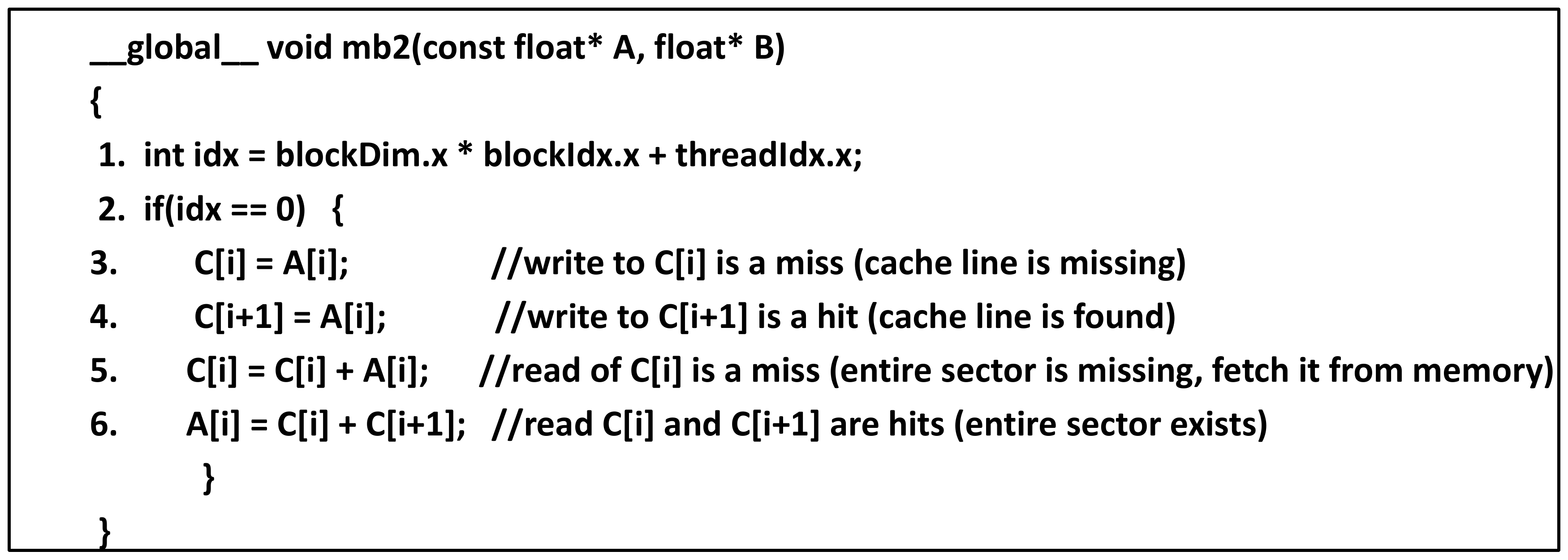}
    \caption{Microbenchmark to explore the Volta L2 cache write policy \label{fig:demystiying_l2write}.}
\end{figure}

Figure~\ref{fig:coal_strides} plots the number of L1 cache reads
and writes per warp.
In figure~\ref{fig:demystiying_l2write},
when stride=1, each thread reads a different cache
line, resulting in 32 accesses for each warp.
When the stride=32, the memory access is converged,
and all the threads within the same warp
will access the same cache line,
however we receive four read accesses at L1 cache,
as shown in figure.
This means that the access granularity of the L1 cache is 32B.
To determine if the cache has a true 32B line size or if
the line size is still 128B (the cache line size in
GPGPU-Sim's modeled Fermi coalescer is 128B), with 32B 
sectors~\cite{liptay2000structural,seznec1994decoupled}, we created
an additional microbenchmark. Our line-size-determining microbenchmark
fills the L1 cache, evicts one entry then attempts to re-access
the data loaded into cache. We find that the eviction granularity
of the cache is 128B, indicating that the L1 cache has 128B lines with
32B sectors.
Furthermore, the coalescer operates across eight threads, i.e. the 
coalescer tries to coalesce each group of eight threads separately to
generate sectored accesses.
We ran these same microbenchmarks with L1 bypassing
turned on and we observed similar behavior at the L2 for both reads
and writes, indicating that the L2 cache has a similar configuration.

\subsection{L2 cache \label{sec:memory-l2}}

Since high throughput memory systems are especially sensitive
to write traffic, the handling of writes at the L2 is very important.
The existing GPGPU-Sim memory model fetches 128B cache lines
from DRAM on write misses at the L2 and uses an allocate-on-fill
writeback policy.
To determine what the behaviour of writes at the L2 is in
real hardware, we created the small microbenchmark in Figure~\ref{fig:demystiying_l2write}.
The microbenchmark confirmed that the L2 is writeback with a 
write-allocate policy.
However, we observed some interesting, previously unmodeled behaviour
that occurs on write misses.
In literature, there are two different write allocation policies~\cite{jouppi1993cache},
\textit{fetch-on-write} and \textit{write-validate}.
In fetch-on-write, when we write to a single byte of a sector,
the L2 fetches the whole sector then merges the written portion to
the sector and sets the sector as modified.
In the write-validate policy, no read fetch is required,
instead each sector has a bit-wise write-mask.
When a write to a single byte is received,
it writes the byte to the sector, sets the corresponding write bit and 
sets the sector as valid and modified.
When a modified cache line is evicted, the cache line is written back to
the memory along with the write mask. It is important to note that,
in a write-validate policy,
it assumes the read and write granularity can be in terms of bytes
in order to exploit the benefits of the write-mask.
In fact, based on our micro-benchmark shown in 
Figure~\ref{fig:demystiying_l2write},
we have observed that the L2 cache applies something similar to
write-validate.
However, all the reads received by L2 caches from the coalescer are
32-byte sectored accesses. Thus, the read access granularity (32 bytes)
is different from the write access granularity (one byte).
To handle this, the L2 cache applies a different write allocation
policy, which we named \textit{lazy fetch-on-read}, that is a compromise
between write-validate and fetch-on-write.
When a sector read request is received to a modified sector,
it first checks if the sector write-mask is complete,
i.e. all the bytes have been written to and the line is fully readable.
If so, it reads the sector, otherwise, similar to fetch-on-write,
it generates a read request for this sector and merges it with the
modified bytes.
In Figure~\ref{fig:demystiying_l2write}, the kernel runs for a single
thread, and we inserted some dummy code (not shown in figure) to prevent
compiler from register allocating the accesses.
In Figure~\ref{fig:demystiying_l2write},
a four-byte float is written at line 3.
When it reads it back at line 5, it is a miss. 
This is because the sector is not full yet.
Note that, in a write-validate policy, this should be a hit. 
Later, at line 6, reading the same four bytes and the
next four bytes are both hits, because the whole sector is fetched from the read to C[i] at line 5.

\subsection{Streaming Throughput-oriented L1 cache \label{sec:memory-l1}}

\begin{figure}
    \centering
    \includegraphics[width=85mm]{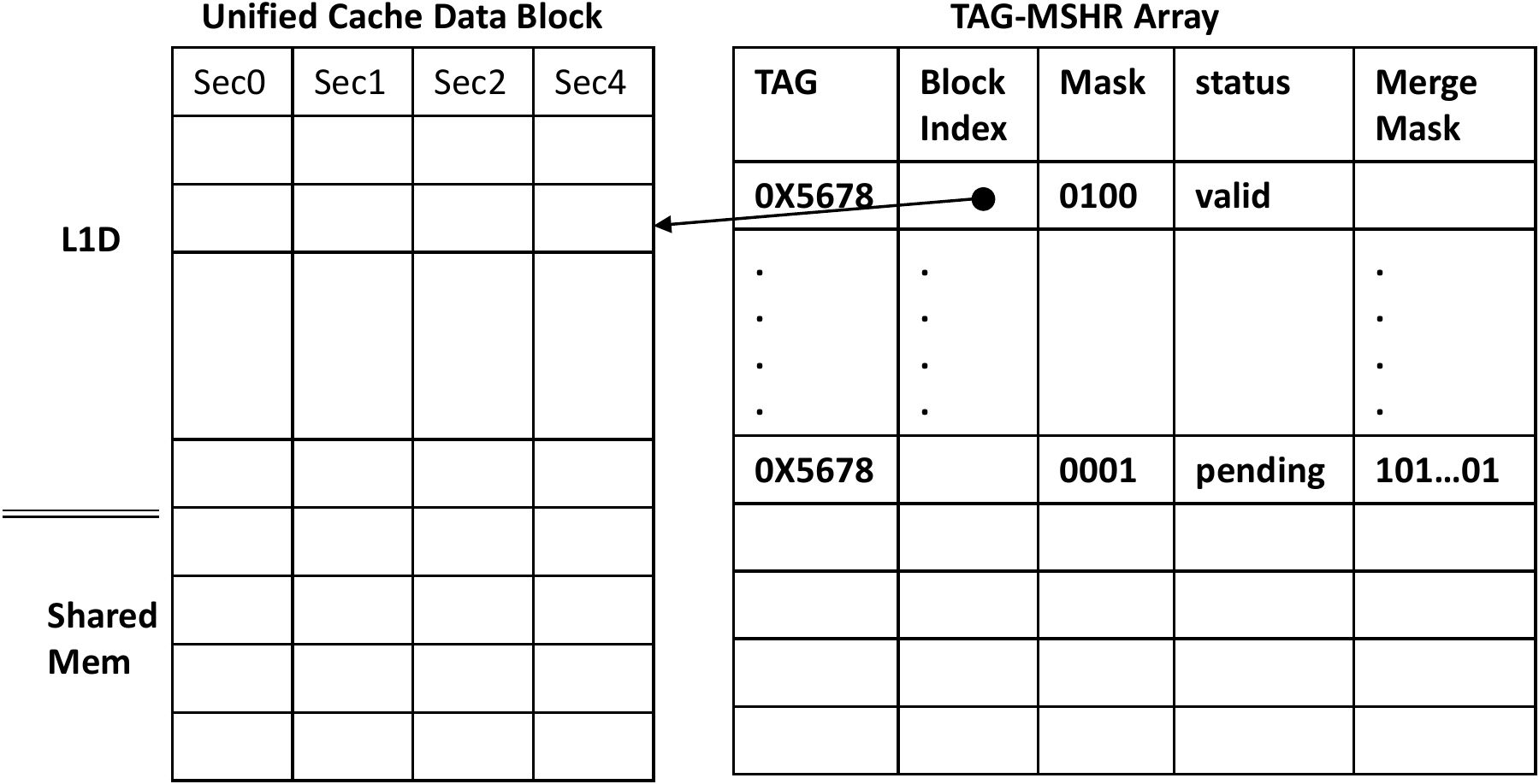}
    \caption{Our Assumed Organization for Streaming L1 Cache in Volta \label{fig:demystiying_streaming}.}
\end{figure}

The L1 cache in Volta is what NVIDIA is calling a {\em streaming cache}~\cite{voltacache}. 
It is streaming because the documentation states that it allows unlimited cache misses to be
in flight regardless the number of cache lines per cache set~\cite{v100micro}.
In other words, L1 cache throughput is not limited by cache resources (e.g, miss-status holding
registers (MSHRs)~\cite{tuck2006scalable} and cache lines). To achieve this, we change the
miss allocation policy in the L1 to be \textit{ON\_FILL} rather than \textit{ON\_MISS} to elimante
line allocaiton fails.
Further, based on our analysis from running some micro-benchmarks, 
imilar to ~\cite{nugteren2014detailed}, we find the number of MSHRs in Volta has increased
substantially compared to previous generations.
We observe that, with just two SMs running a streaming workload, Volta can fully utilize the memory system.
Also, we notice that, independent of the number of L1 configured size, the number of MSHRs available are the same,
even if more of the on-chip SRAM storage is devoted to shared memory.
Based on that, Figure~\ref{fig:demystiying_streaming} depicts our assumed organization of streaming cache.
We believe that unified cache is a plain SRAM where sectored data blocks are shared between the L1D and the CUDA
shared memory. It can be configured adaptively by the driver as we discussed earlier.
We assume that the L1D's TAG and MSHR merging functionality are combined together in a
separate table structure (TAG-MSHR table). Since, the filling policy is now \textit{ON\_FILL},
we can have more TAG entries and outstanding requests than the assigned L1D cache lines.
When a new 32B sector request generated from the coalescer is received,
it checks the TAG array. If it hits and the status is valid, it uses a block index to access the data array.
If it is a hit to a reserved sector (i.e. the status is pending),
it sets its corresponding warp bit in the merging mask (64 bits for 64 warps).
When the pending request comes back, it allocates a cache line/sector in the data block
and sets the allocated block index in the table. Then, the merged warps access the
sector, on a cycle-by-cyle basis.

%% file: correlation.tex
\section{Hardware Correlation}
\label{sec:hw-correl}

Throughout this section we will study the correlation of both the older, less detailed
GPGPU-Sim model and our enhanced version versus a silicon NVIDIA Volta TITAN V card.
The correlation figures presented throughout this section plot the collected hardware
number on the  x-axis and the GPGPU-Sim number on the y-axis.
Depending on the statistic, there are up to  \textasciitilde2400 datapoints (\~1200 kernels
with one data point for each of the old and new models) on each graph. 
The center of the larger orange circle represent the new GPU model and the small black
dot is the old model.
The {\em New Model} represents the results after performing the memory system
enhancements detailed in Section~\ref{sec:exp}, while the {\em Old Model} is our best-effort
attempt to model the Volta card using the existing codebase with only configuration file changes
to the existing GPGPU-Sim memory model. We note that both configurations make use of
the virtual PTX ISA, which is different from the SASS machine code used in the Volta.
Using PTX enables us to run many more workloads than with the limited support in GPGPU-Sim
for PTXPLUS. Moreover, recent work~\cite{jain2018quantitative} has demonstrated that while using PTXPLUS
can slightly improve the correlation of compute-intensive workloads, but has largely no effect on
memory-intensive applications and that error in the memory system is the most pervasive cause of error
in the simulator, independent of ISA choice.

This section begins by presenting the improvement in overall performance correlation
that results from our enhanced memory model. Then we detail how well the performance
counters in the memory system correlate to both the new and old models.

\subsection{Overall Cycle Correlation}

\begin{figure}
    \centering
    \includegraphics[width=85mm]{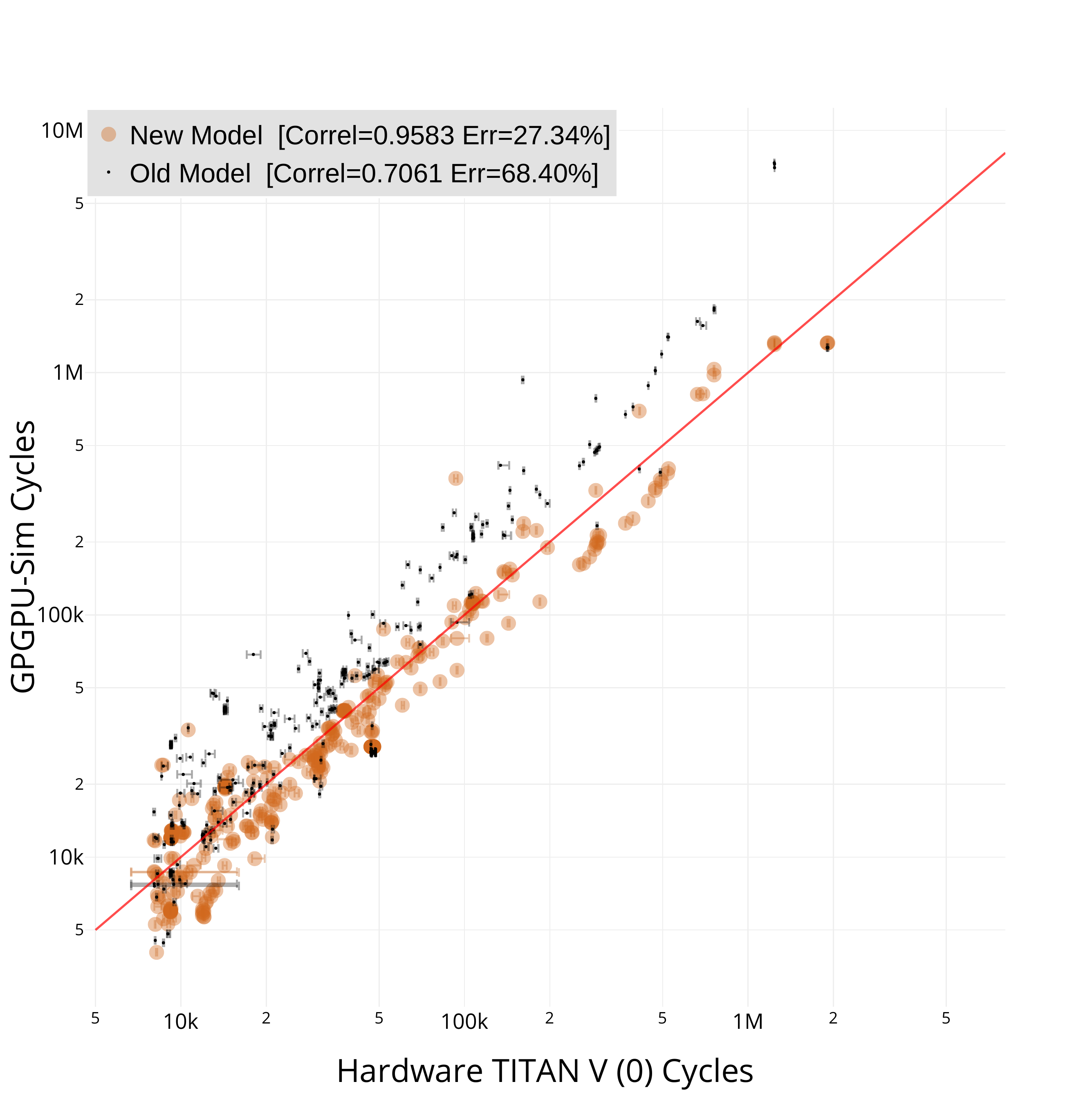}
    \caption{Execution time correlation for the NVIDIA TITANV\label{fig:cycles-correl}. }
\end{figure}

Figure~\ref{fig:cycles-correl} plots the resulting error and correlation of
cycles in both the old and new models. The error bars on the x-axis represent the
range of the silicon GPU execution time for this particular kernel over
10 successive runs of the application. To minimize the effect of
noisy, small kernels on the correlation results, only kernels that ran for at least 8000
GPU cycles in hardware are considered for cycles correlation.
Both the correlation mean absolute error of cycles are significantly
better using the new memory model. In particular, the mean absolute error is reduced by 
~$2.5\times$. This result confirms our working assumption that increasing the level of
detail in the memory system would decrease the overall error in execution time.

\subsection{Correlating the L1}

By default the L1 cache is enabled in Volta (unlike in prior generations like Pascal, where the L1 cache was
shared with the texture and disabled by default).
Therefore; to run all of our test, we assume that the L1 is enabled and our newly implemented
adaptive L1/shared memory partitioning algorithm determines it's size. Independent of the number of
data lines allocated, we assume the streaming cache is still in use and that the number of tags are
available the same, even if more of the on-chip SRAM storage is devoted to shared memory.

\begin{figure}
    \centering
    \includegraphics[width=85mm]{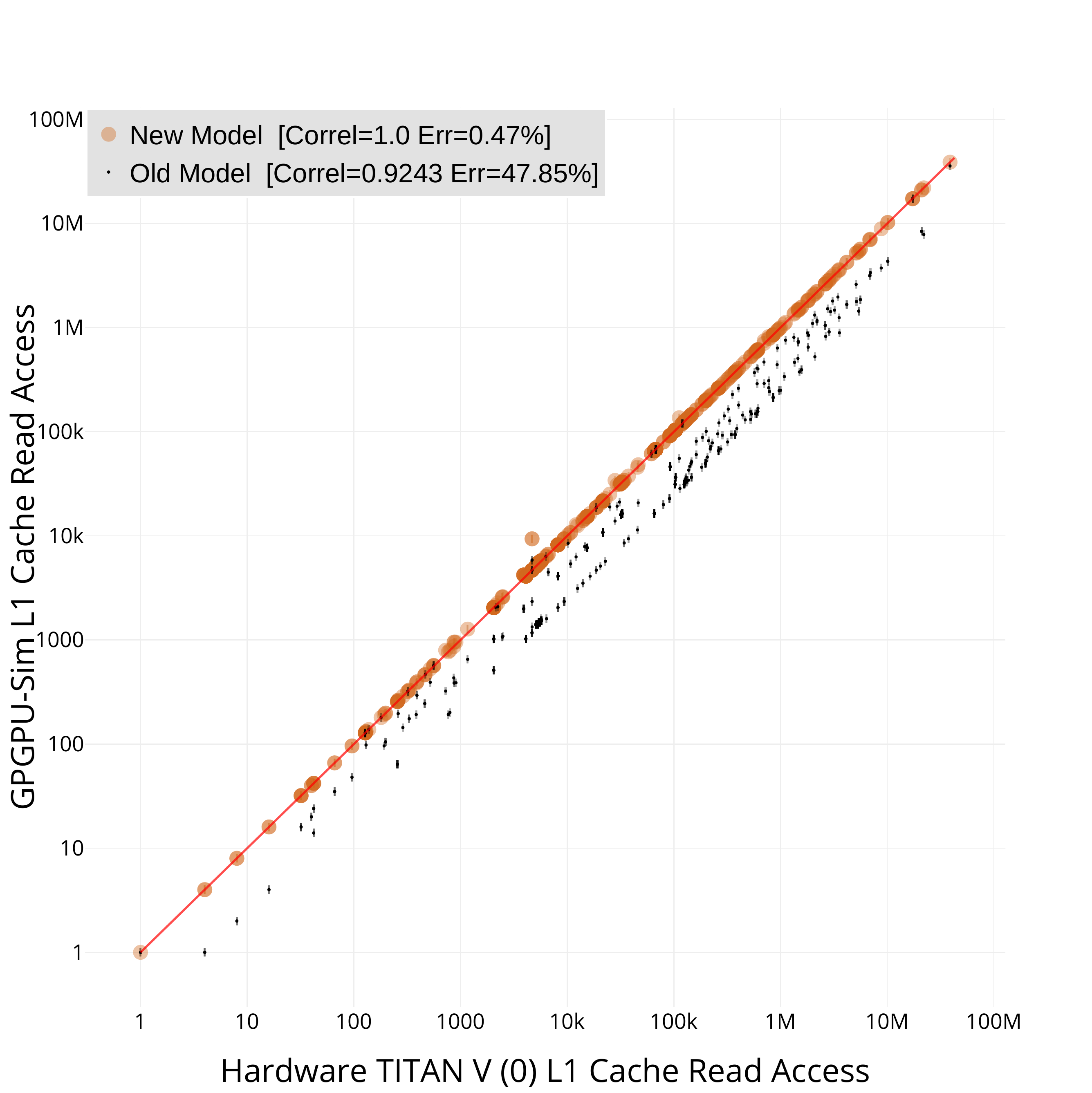}
    \caption{ L1 read accesses correlation for the NVIDIA TITANV\label{fig:l1access-correl}. }
\end{figure}

Figure~\ref{fig:l1access-correl} plots correlation for the L1 read accesses
generated. Note that for all the counter correlation graphs, there are no errors on the hardware
measurement, as the hardware nvprof profiler runs the app several times to collect the statistics and
we have found these to be much more stable than execution time over successive runs of the apps.
Moreover, we plot the results for all the kernels run, regardless of their execution time,
as the counters are less noisy and more consistent than execution time.
As figure~\ref{fig:l1access-correl} shows, the error in the new model has been reduced to almost zero,
where the old model gives almost 50\%. This is primarily due to the implementation of
Volta coalescer and the fact that the L1 cache is sectored. The consistent band of accesses in the old model
at $y = 4x$ are reflective of this. 1 128B access in the old model will count as 4 sector
accesses in the new model. Completely divergent applications show consistent behaviour between both
models since 32 unique sectors are generated in the new model and 32 unique cache lines are generated
in the old model.

\begin{figure}
    \centering
    \includegraphics[width=85mm]{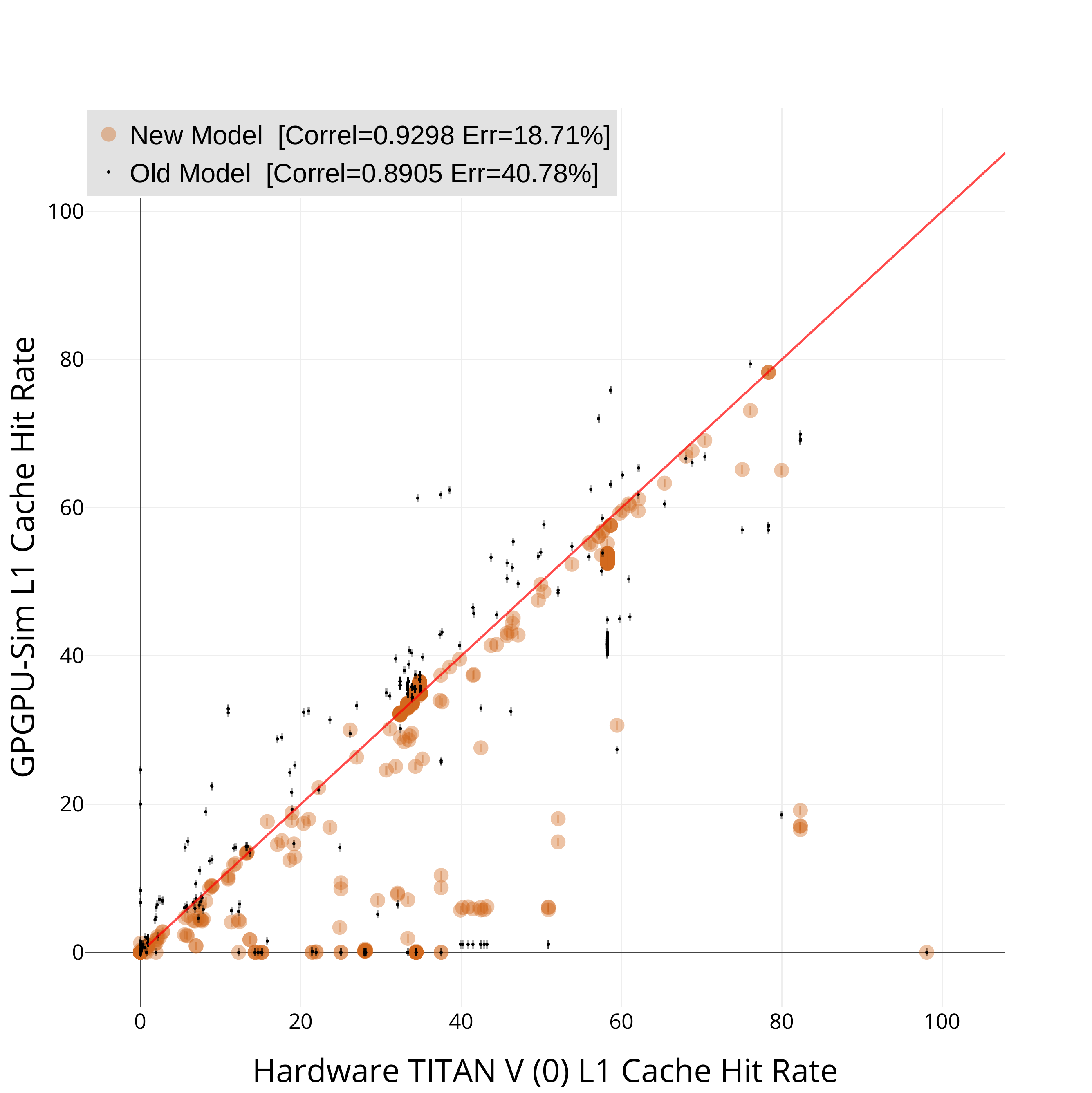}
    \caption{ L1 hit rate correlation for the NVIDIA TITANV\label{fig:l1hitrate-correl}. }
\end{figure}

Figure~\ref{fig:l1hitrate-correl} plots the corresponding hit rate for the L1 cache.
We note that this statistic is particularly hard to correlate, given that the warp scheduling
policy, cache replacement policy and hit/miss accounting decisions in the profiler can skew the results.
For example, the through rigorous microbenchmarking we determined that the hardware profiler
will count a sector miss on a cache line whose tag is already present as a "hit", but will still
send the sector access down to the L2. Effectively, the profiler appears to be counting 128B
line cache tag hits, even if the sector is still fetched and the instruction must wait.
Even with these variables, our improved cache model and coalescer achieve only 20\% error
(a 2$\times$ improvement over the old model) and a correlation of 93\%.
In the apps where pervasive error still exist, we believe more detailed
reverse engineering of the warp scheduling policy and L1 eviction policy would help.

\subsection{Correlating the L2}
\label{sec:l2-corr}

\begin{figure}
    \centering
    \includegraphics[width=85mm]{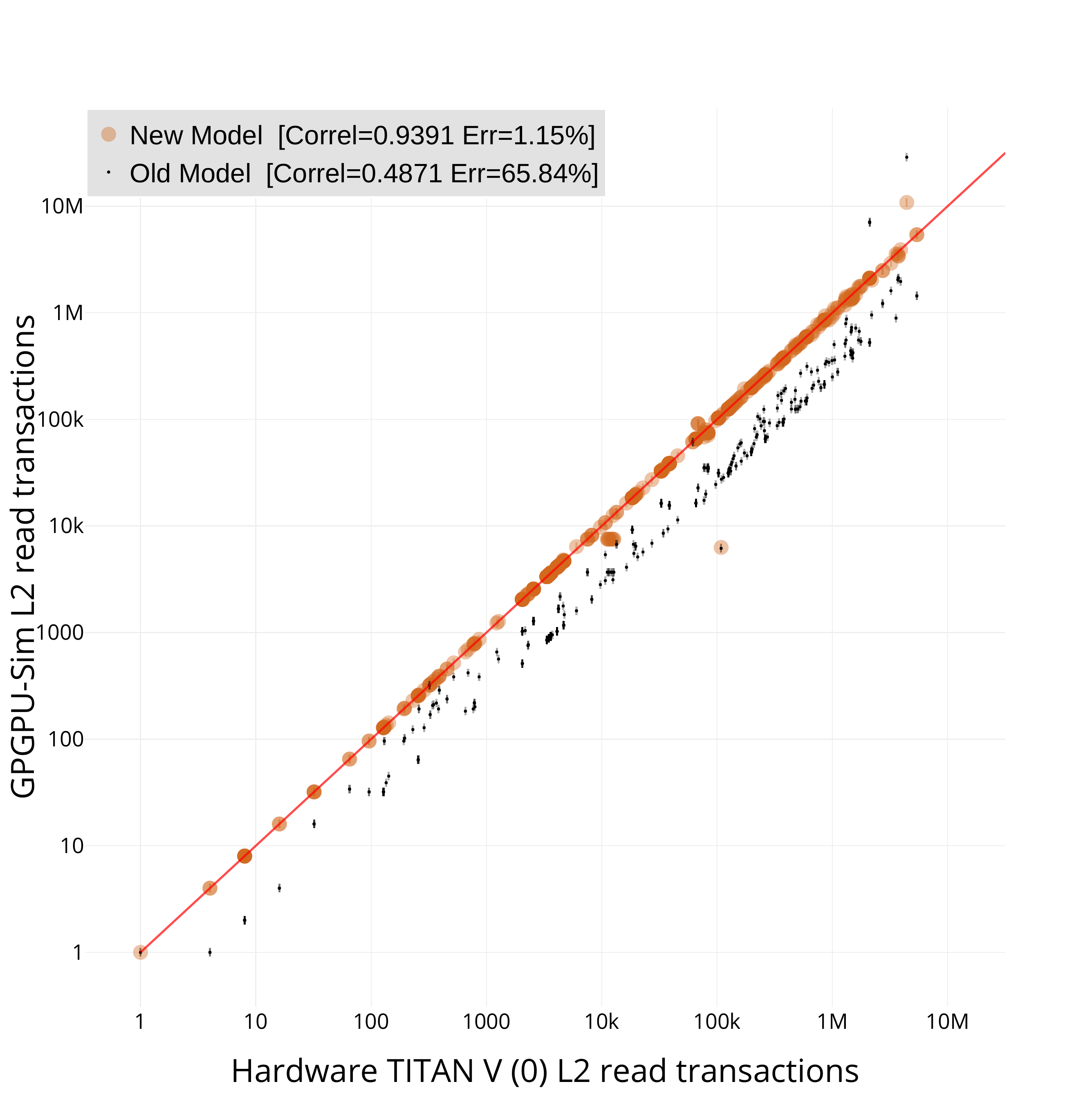}
    \caption{L2 read transaction correlation with NVIDIA TITAN V\label{fig:l2-read-correl}. }
\end{figure}

To determine how well the memory system is operating at the memory controller
level, we begin by examining the performance of the memory-side L2 data cache.
Figure~\ref{fig:l2-read-correl} plots the correlation of L2 read accesses versus hardware.
Both correlation and error are much improved with the new model, with absolute error
reaching only 1.15\%. Interestingly, the relatively high
error in L1 miss rate does not end up resulting in error at the L2 level.
This leaves us to believe that most of the error is a result of accounting peculiarities
for L1 cache hits, like the sector miss situation described previously.

In terms of improvement over the old model, there is a significant fraction of
error coming from accessing 128B lines instead of 32B sectors from the L1,
again note the old model line at $y=4x$.
We also collected information on write transactions, and noticed similar behaviour
which is summarized in Table~\ref{ta:corr}.

\begin{figure}
    \centering
    \includegraphics[width=85mm]{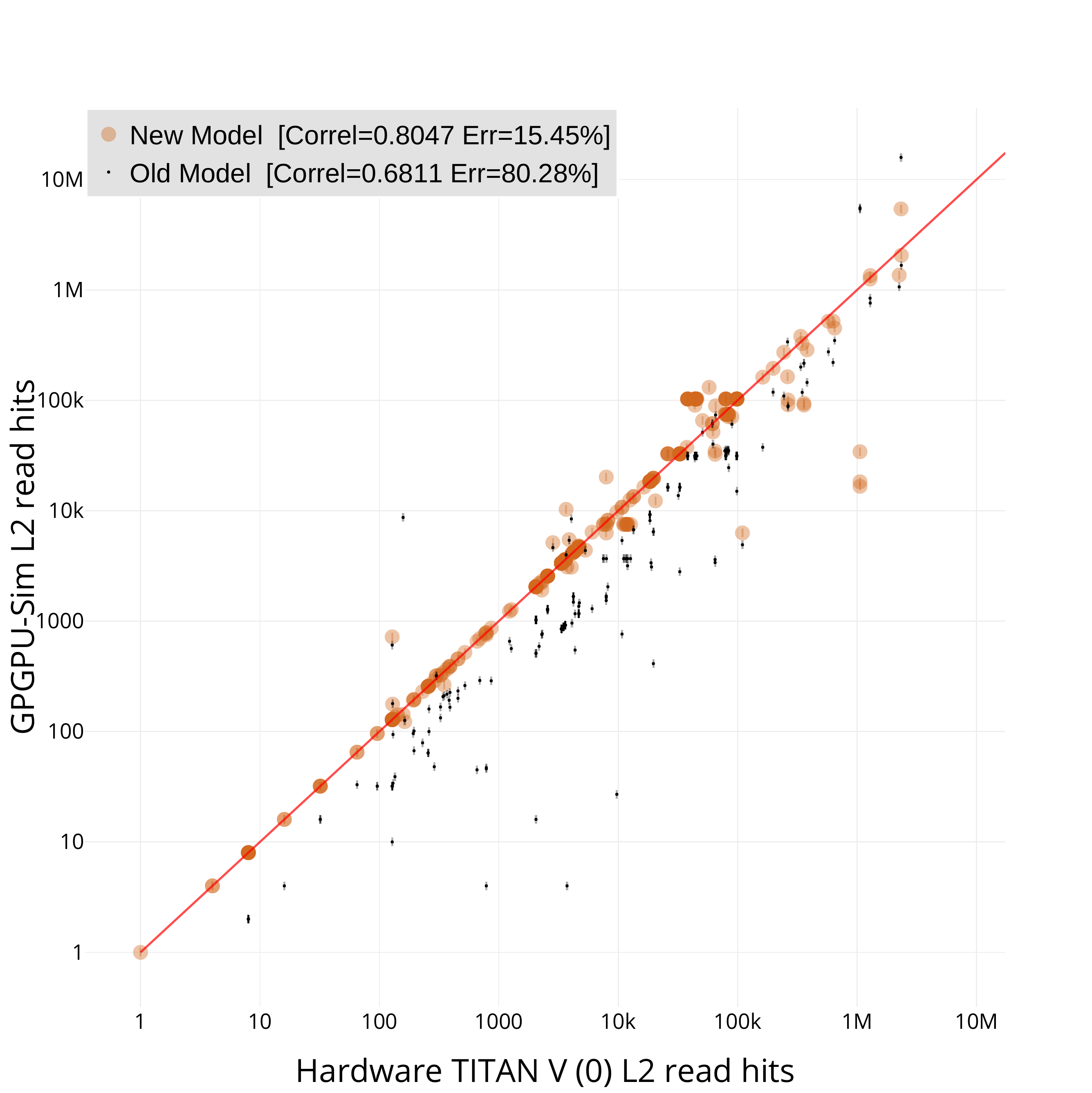}
    \caption{L2 read hit transaction correlation with NVIDIA TITAN V\label{fig:l2-read-hit-correl}. }
\end{figure}

We also attempted to correlate the L2 cache read hit rate with simulation results,
but found that the profiler gives inconsistent results, some of which are greater than 100\%.
So instead, we correlated with the number of L2 hits, which gave more grounded readings.
Figure~\ref{fig:l2-read-correl} presents the L2 read hit correlation.
Overall, the hit error is roughly the same as the error on the L1 cache, \~15\%.
Again, at the L2 scheduling effects from the order in which memory accesses reach the L2 will have an
effect on it's hit rate. The older model suffers significant error at the L2, even more than at
the L1 (80\% versus 40\%). Our model shows significantly better correlation for two primary several reasons.
First, since the number of accesses is more correlated the overall behaviour should be more similar.
Second, our model fills the L2 on memory copies from the CPU avoiding many unnecessary misses in the L2
when kernels with smaller data sets begin. The majority of excessively low old model points are
examples of this. We also collected data on the number of read hits, which showed a similar trend.

\begin{figure}
    \centering
    \includegraphics[width=85mm]{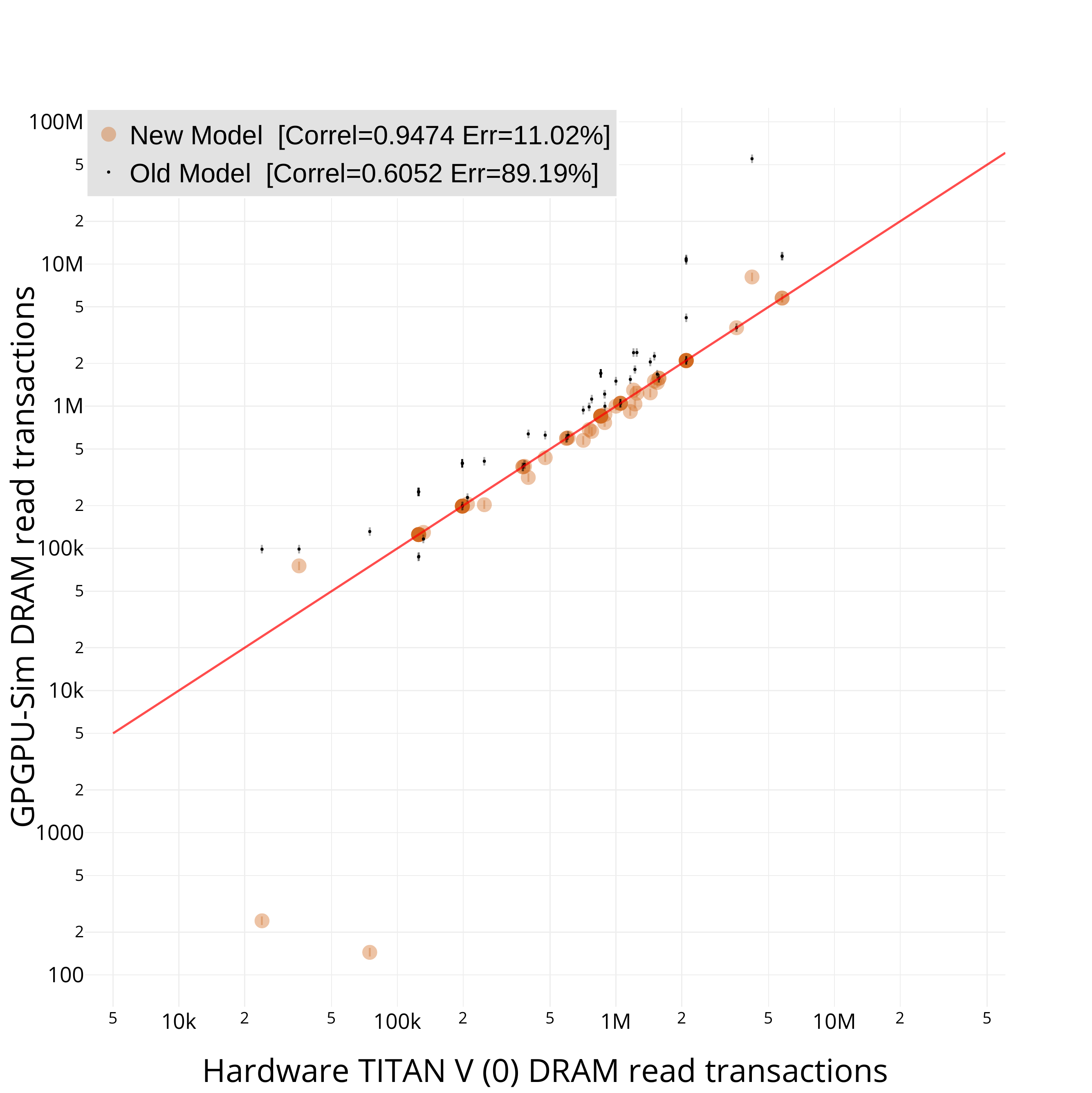}
    \caption{DRAM read transactions correlation versus NVIDIA TITAN V\label{fig:dram-reads}.}
\end{figure}

\subsection{Correlating DRAM}

Finally, Figure~\ref{fig:dram-reads} plots the correlation of DRAM read accesses.
The error rate here is 15\% (v.s. 89\% for the old model). As expected, small errors in the upper levels
of memory propagate into larger errors at the lower levels. The new model DRAM read correlation
remains high and is again more accurate for larger kernels. Interestingly, the reason for the massive error
in the old model was dues to its per-cacheline fetch on write policy
(detailed in Section~\ref{sec:memory-l2}). The end result of that policy was that every write to the
writeback L2 fetched 4 32-byte memory values from DRAM, which is why the DRAM reads of
the old model consistently are consistently overestimated. Our updated model has eliminated this problem
by sectoring the cache and implementing a more intelligent write policy, explained in Section~\ref{sec:memory-l2}.
We note that we only collect data here for kernels that have at least 1000 dram accesses, as we found the
counter values from hardware tended to be inconsistent at such small numbers and both the new and old
model resulted in massive amounts of error due to being off by relatively few memory accesses.



%% file: motivation.tex
\section{Design decision case study \label{case-study}}

\begin{figure}
    \centering
    \includegraphics[width=85mm]{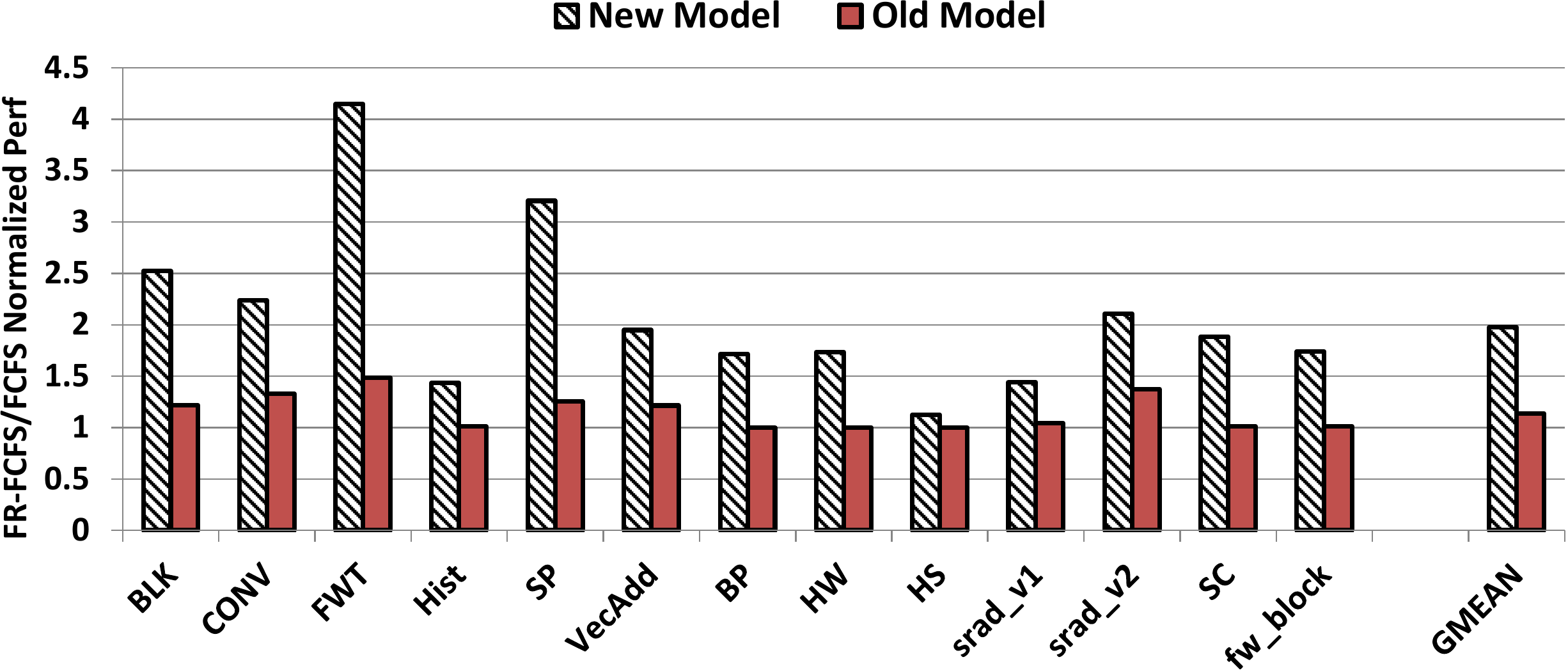}
    \caption{FR\_FCFS performance normalized to the FCFS in both old and new model.\label{fig:dram-shced} }
\end{figure}

This section presents the effect of inaccurate memory system modeling
on architectural design decisions. An inaccurate architecture model
may lead to  unrealistic issues or incorrect conclusions that are
not relevant to state-of-the-art designs already being used in industry.
This section demonstrates how our more accurate GPU model removes the well-studied
bottleneck of L1 cache throughput and opens up new research opportunities
in more advanced memory access scheduling.

\textbf{DRAM scheduler}: Throughput-oriented GPGPU workloads are
often sensitive to memory bandwidth.
Thus, the DRAM memory scheduler plays an important role to efficiently utilize the memory bandwidth
and improve the performance of memory-sensitive workloads.
Figure \ref{fig:dram-shced} shows the sensitivity of GPGPU workloads for
two memory scheduling policies: the na\"ive first-come first-serve scheduling (FCFS)
and the advanced out-of-order First-row-ready first-come first-serve scheduling
(FR\_FCFS) \cite{rixner2000memory}.
The data is shown for FR\_FCFS performance normalized to the FCFS in both old and new model.
As shown in figure, the memory scheduling policy seems to be more important in the new model
than the old model. In the old model, some workloads are insensitive or show little difference between the two
scheduling policies. On average, 20\% performance increase was observed when applying FR\_FCFS in the old model. 
Yet, in the new model, the memory scheduling policy seems to be more critical. Applying FR\_FCFS in the new model improves the performance by 2X on average. This is due to updating the model to use the NVIDIA Volta memory access coalescing rules and modeling  sectored and advanced write allocation policies in the L2 cache. 
Further, with the new features and capabilities of the GDDR5X and HBM such
as dual-bus interface \cite{o2014highlights}, pseudo-independent accesses\cite{gddr5xstandard2013high}
and per-bank refresh command \cite{hbmstandard2013high},
memory scheduling will become a more critical issue to investigate.
This experiment demonstrates how accurately modeling contemporary GPU hardware reveals performance issues obscured using a less accurate simulation.

\begin{figure}
    \centering
    \includegraphics[width=85mm]{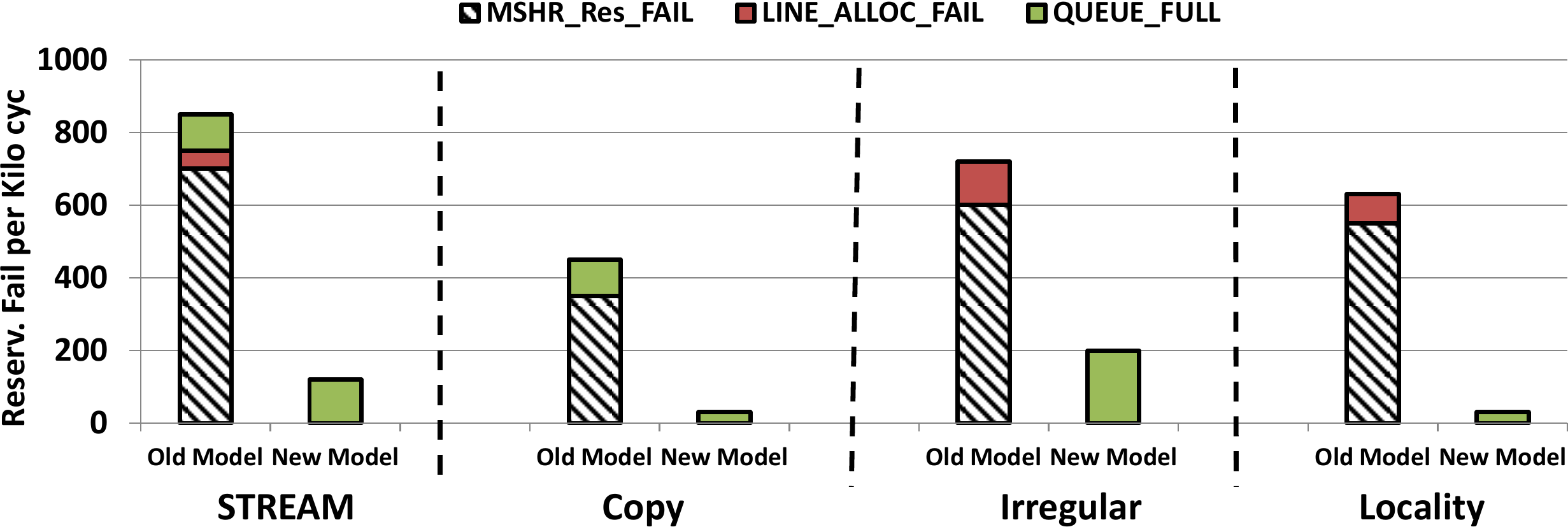}
    \caption{L1 cache reservation fails per kilo cycles for some microbenchmarks.\label{fig:l1-cache-contention} }
\end{figure}

\begin{figure}
    \centering
    \includegraphics[width=85mm]{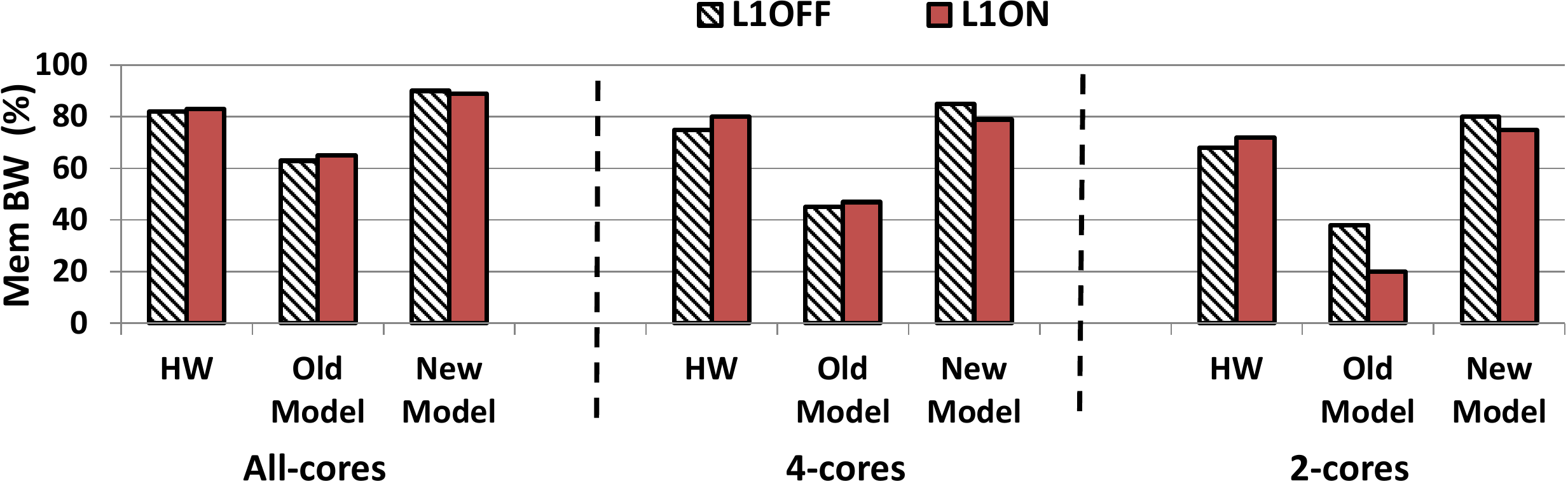}
    \caption{BW utilization of STREAM workload for TITANV HW, Old Model and New Model when L1 cache is turned On and Off\label{fig:l1-cache-throughput}. }
\end{figure}

\textbf{L1 cache throughput}: The L1 cache throughput bottleneck for GPUs has been well studied in literature \cite{jia2014mrpb,sethia2015mascar,kim2017packet,koo2017access}. Due to massive multithreading,
GPGPU L1 caches can suffer from severe resource contention
(e.g. miss status handling registers (MSHRs) and cache line allocation~\cite{jia2014mrpb}).
However, contemporary GPUs are designed such that the L1 cache is not a throughput bottleneck.
The new GPUs employ advanced architecture techniques to improve L1 cache throughput,
such as sectored, banked and streaming L1 caches (as discussed in section \ref{sec:memory-l1}).
The new L1 cache eliminates the excessive L1 cache throughput contention.
Figure \ref{fig:l1-cache-contention} depicts the L1 cache reservation fails
per kilo cycles for the new vs old model.
The new L1 model eliminates reservation fails and the
L1 cache is no longer a throughput bottleneck.
Figure \ref{fig:l1-cache-throughput} demonstrates the bandwidth utilization when we 
run a highly streaming application (STREAM) on the new and old models vs the hardware.
The figure plots results with the L1 cache on and off as we scale down the number of active cores,
while keeping memory bandwidth the same.
As shown in figure, the hardware is able to achieve 82\%, 75\% and 68\%
of the theoretical bandwidth when we run the workload on all SM cores, 4 and 2 cores respectively.
The old model is not able to achieve the same bandwidth utilization due to the L1 cache bottleneck
and inaccurate L2 cache replacement policy (as discussed in section \ref{sec:memory-l2}).
Furthermore, the L1 cache is a clear bottleneck in the old model, as turning the cache on severely
decreases memory utilization. Turning the L1 cache on in the new model has no effect on bandwidth
utilization and very closely matches the performance of the hardware.

%% file: related.tex
\section{Related Work }
\label{sec:related}

There are quite a few GPU simulators and instrumentation tools that are found in literature. 
Xun et al. \cite{gong2017multi2sim} model GPU Kepler architecture \cite{Kepler} that is integrated with Multi2sim simulation infrastructure \cite{ubal2012multi2sim}. The new simulator was validated using few CUDA SDK applications. Barrio et al.\cite{del2006attila} propose ATTILA which is a trace-driven GPU simulator that models the old GT200 architecture. Raghuraman et al. \cite{balasubramanian2015miaow} propose Miaow, an open source RTL implementation of the AMD Southern Islands GPGPU ISA. Jason et al. \cite{industry28} propose gem-gpu simulator in which they integrate GPGPU-sim with the well-known CPU simulator GEM5. The GPGPU-Sim memory system was replaced with GEM5 memory infrastructure.  Beckmann~\cite{beckmann2015amd,gutierrez2018lost} proposed gem5’s GPU, which models AMD GCN architecture. Kim et al.~\cite{kim2012macsim} proposed Mac-sim a trace-driven fermi-based GPU simulator based on Ocelot~\cite{diamos2010ocelot}. Stephenson et al.\cite{stephenson2015flexible} introduce SASSI, a low-level assembly-language instrumentation tool for GPUs. Numerous works tried to demystify GPU architecture and memory system details through micro-benchmarks~\cite{wong2010demystifying,nugteren2014detailed,mei2017dissecting,voltacitadel}. 

Previous works have studied the deficiencies of GPGPU-Sim infrastructure. Nowatzki et al.\cite{nowatzki2014gem5} show that GPGPU-Sim is very inaccurate in modeling some in-core parameters and connection delays. Hongwen et al.\cite{daidemand} argue that GPGPU-Sim has a very weak L1 cache model and they suggest few enhancement to improve the L1 cache throughput. Also, Nugteren\cite{nugteren2014detailed} validate GPGPU-Sim against real Fermi hardware and find that the simulator does not accurately model the L1 cache throughput, especially the  number of miss status holding registers. There have been many works in literature that claim they model Kepler, Maxwell and Pascal architectures by scaling the Fermi resource, however none of these works correlate their new model against real hardware. A recent work \cite{jain2018quantitative} shows that there is a huge gap in GPGPU-Sim memory system modeling when correlated against Pascal hardware. 

Our new simulator demystifies and models the latest Volta architecture and shows a high correlation with real hardware, especially in memory system, on a wide range GPGPU workloads. To the best of our knowledge, this is the first work that accurately models a GPU Volta architecture in details. There are other few works\cite{chatterjee2017architecting,o2017fine} that claim they use in-house closed-source modern Pascal simulator in their experiments, however, our new simulator is open-source.





%% file: conclusion.tex
\section{Conclusion }
\label{sec:conclusion}
This paper presents the most accurate open-source
model of a contemporary GPU to date.
Our detailed hardware characterization
and modeling effort decreases the execution time error
in GPGPU-Sim, the most popular GPU simulator in use today,
by $2.5\times$ when running
a comprehensive suite of contemporary workloads.
Through our detailed micro-benchmarking and remodeling, we have uncovered a number of
important design decisions made in the GPU memory system that no
previous work has identified and no open-source simulator has attempted to model.

The paper goes on to concretely demonstrates the effect this memory system error
has on design decisions.
Specifically, we show that error in the less detailed memory system model
discounts the performance effects of out-of-order memory
access scheduling, hampering potential research on the topic.
Conversely, the less detailed model over-estimates the
impact of L1 cache contention, as the throughput of the L1 cache is no longer
a bottleneck to the memory system in contemporary GPUs.

In the process of creating our detailed model, we created the {\em Correlator}
simulation and correlation infrastructure that can be easily used to
validate new modeling efforts as new hardware and simulation models are created.
We hope that releasing all the infrastructure used to perform this validation,
along with the more detailed model itself, will help facilitate the continued development of
simulation models that keep up with changes in contemporary
industrial designs so that academics can more accurately study what should
comes next.


